\documentclass[prd,twocolumn,nofootinbib]{revtex4}

\usepackage[letterpaper,hmargin=1in,vmargin=1in]{geometry}

\usepackage{graphicx,epstopdf,amsmath,amsfonts,amssymb,color}

\def\be{\begin{equation}}
\def\ee{\end{equation}}
\def\ba{\begin{eqnarray}}
\def\ea{\end{eqnarray}}
\def\ge{\mathrel{\raise.3ex\hbox{$>$\kern-.75em\lower1ex\hbox{$\sim$}}}}
\def\la{\mathrel{\raise.3ex\hbox{$<$\kern-.75em\lower1ex\hbox{$\sim$}}}}

\def\simgt{\mathrel{\raise.3ex\hbox{$>$\kern-.75em\lower1ex\hbox{$\sim$}}}}
\def\simlt{\mathrel{\raise.3ex\hbox{$<$\kern-.75em\lower1ex\hbox{$\sim$}}}}

\newcommand{\bi}[1]{\bibitem{#1}}
\newcommand{\fr}[2]{\frac{#1}{#2}}

\newcommand{\nc}{\newcommand}

\nc{\gone}{\bar g_{\pi NN}^{(1)}}
\nc{\gzero}{\bar g_{\pi NN}^{(0)}}
\nc{\al}{\alpha}
\nc{\ga}{\gamma}
\nc{\de}{\delta}
\nc{\ep}{\epsilon}
\nc{\ze}{\zeta}
\nc{\et}{\eta}
\nc{\ka}{\kappa}
\nc{\rh}{\rho}
\nc{\si}{\sigma}
\nc{\ta}{\tau}
\nc{\up}{\upsilon}
\nc{\ph}{\phi}
\nc{\ch}{\chi}
\nc{\ps}{\psi}
\nc{\om}{\omega}
\nc{\Ga}{\Gamma}
\nc{\De}{\Delta}
\nc{\La}{\Lambda}
\nc{\Si}{\Sigma}
\nc{\Up}{\Upsilon}
\nc{\Ph}{\Phi}
\nc{\Ps}{\Psi}
\nc{\Om}{\Omega}
\nc{\ptl}{\partial}
\nc{\del}{\nabla}
\nc{\ov}{\overline}
\nc{\newcaption}[1]{\centerline{\parbox{15cm}{\caption{#1}}}}
\nc{\us}{U(1)$_S$}

\def\beq{\begin{equation}}
\def\eeq{\end{equation}}
\def\bmat{\begin{displaymath}}
\def\emat{\end{displaymath}}
\def\bear{\begin{eqnarray}}
\def\eear{\end{eqnarray}}
\def\ba{\begin{eqnarray}}
\def\ea{\end{eqnarray}}
\def\bery{\begin{array}}
\def\ery{\end{array}}
\def\bit{\begin{itemize}}
\def\eit{\end{itemize}}
\def\ben{\begin{enumerate}}
\def\een{\end{enumerate}}
\def\btab{\begin{tabular}}
\def\etab{\end{tabular}}
\def\btbl{\begin{table}}
\def\etbl{\end{table}}
\def\bfig{\begin{figure}[htb]}
\def\efig{\end{figure}}
\def\bpic{\begin{picture}}
\def\epic{\end{picture}}

\def\ga{\mathrel{\raise.3ex\hbox{$>$\kern-.75em\lower1ex\hbox{$\sim$}}}}
\def\la{\mathrel{\raise.3ex\hbox{$<$\kern-.75em\lower1ex\hbox{$\sim$}}}}
\def\gappeq{\mathrel{\rlap {\raise.5ex\hbox{$>$}}
{\lower.5ex\hbox{$\sim$}}}}
\def\lappeq{\mathrel{\rlap{\raise.5ex\hbox{$<$}}
{\lower.5ex\hbox{$\sim$}}}}

\def\gyr{{\rm \, G\kern-0.125em yr}}
\def\mev{{\rm \, Me\kern-0.125em V}}
\def\gev{{\rm \, Ge\kern-0.125em V}}
\def\tev{{\rm \, Te\kern-0.125em V}}

\begin{document}
 
\title{Higgs decays to dark matter: beyond the minimal model}

\author{Maxim Pospelov$^{\,1,2}$ and Adam Ritz$^{\,1}$}
\affiliation{$^{\,1}$Department of Physics and Astronomy, University of Victoria, 
     Victoria, BC, V8P 5C2 Canada\\
     $^{\,2}$Perimeter Institute for Theoretical Physics, Waterloo,
ON, N2J 2W9, Canada}

\date{September 2011}

\begin{abstract}
\noindent
We examine the interplay between Higgs mediation of dark matter annihilation
and scattering on one hand, and the invisible Higgs decay width on the other, in a 
generic class of models utilizing the Higgs portal. We find that, while the invisible width of
the Higgs to dark matter is now constrained for a minimal singlet 
scalar WIMP by experiments such as XENON100, this conclusion is not robust within more generic examples of 
Higgs mediation. 
We present a survey of simple WIMP scenarios with $m_{\rm DM} < m_h/2$ and Higgs portal mediation, 
where direct detection signatures are suppressed, while the Higgs width is still dominated by
decays to dark matter.
\end{abstract}
\maketitle

\section{Introduction}

The past year has seen impressive progress toward an understanding of electroweak symmetry breaking at the LHC and the Tevatron.
The allowed mass range for its simplest manifestation -- the Standard Model (SM) Higgs boson -- is now limited to 114 - 145 GeV \cite{higgs}, which 
is of course indirectly favored by the global precision electroweak fit. This low mass 
region is notoriously difficult to probe at the LHC, and may also be vulnerable to non-SM decay modes that can hide the Higgs by suppressing
conventional decays even in the simplest extensions of the Standard Model,  see {\em e.g.} \cite{ss,Gunion,NSHiggs}. Thus, while the absence of 
SM-type Higgs decay signatures over the full mass range could be interpreted as evidence in  favor of a nonperturbative mechanism for electroweak symmetry 
breaking, a plausible alternative hypothesis would be that a perturbative Higgs scalar is present but with non-SM decay channels that make 
it hard to identify \cite{ss,Gunion,NSHiggs}.
Among these scenarios, Higgs decays to Dark Matter (DM) through the so-called Higgs portal comprise a distinct and natural 
possibility. While first identified many years ago \cite{SZ},  it is only recently that both Higgs searches \cite{higgs} and dark matter direct detection 
efforts \cite{direct} have reached the level of sensitivity required to make an experimental test of this possibility a reality. 

As the LHC approaches the sensitivity level required to directly observe a SM Higgs in the low mass range, 
underground direct detection experiments probing nuclear recoils of weakly-interacting massive particle (WIMP) dark matter, are also reaching
the important threshold of Higgs-mediated scattering, $\si_{\rm SI} \sim 10^{-45}-10^{-43}$~cm$^2$ \cite{direct}. 
This coincidence naturally focuses attention on the possible interplay between these two probes of the Higgs sector and its interactions.
 In particular, the question arises as to whether a putative invisible Higgs width may be constrained by the presence or absence of any 
direct detection signal. As recently emphasized in the literature for the most economical DM model -- a singlet scalar with Higgs-mediated 
interactions \cite{SZ,McD,BPtV} -- combining the collider limits on a SM-like Higgs with the direct detection
constraints indeed leads to significant restrictions on any  invisible Higgs branching in the low mass $m_h \sim 120$~GeV region \cite{SR,Mambr,He,fhkt}. 
In the present note we will examine the generality of this conclusion in a simple but more generic class of Higgs-mediated  dark matter
interactions, finding that it is far from robust. Indeed, we observe that many scenarios for Higgs mediation in the
dark sector, beyond the minimal singlet scalar, allow for a significant invisible Higgs branching
while being poorly constrained by the results of direct detection experiments.

In order to focus on invisible Higgs decays, we will consider dark matter (and scalar mediators) whose mass is below $m_h/2$, i.e. below 50-60 GeV for
the light Higgs region. For such relatively light states, effective field theory dictates that the largest couplings will be through renormalizable operators, 
and in this case the Higgs portal:
\be
 {\cal L}_{\rm portal} = H^\dagger H \left( A_i S_i + \lambda_{ij} S_i S_j\right), \label{portal}
\ee
where $H$ is the SM Higgs doublet and we allow for at least two real SM-singlet scalar fields $S_i$. These scalars, and 
other components of the hidden sector, could also transform under larger representations of a hidden sector (gauge) group or indeed 
be composite operators, but for simplicity we focus on perturbative singlets as representative of the basic physics involved.  The lightest 
scalar $S$  may be a dark matter candidate itself, or may mediate the interactions of another stable dark matter species in the hidden sector. 
For the latter case, we enumerate the renormalizable possibilities below for a hidden sector fermion $N$,
\be
 {\cal L}_{\rm hid} = (m_N+ \al_i S_i) \bar{N} N + \beta_i S \bar{N} i\gamma_5 N - V(S_i). \label{hidden}
\ee
The potential term may include multiple scalars, and it is assumed that it satisfies the usual requirements 
of vacuum stability. The minimal model contains just one scalar field, and the Higgs portal 
$\lambda_{\rm min}H^\dagger HS^2$ regulates the abundance of $S$ as WIMP DM. It also provides a rigid link between the 
invisible Higgs branching ratio and the DM scattering cross section \cite{SZ,BPtV}. 

Various combinations of the portal and hidden sector couplings $(A,\lambda,\al/\beta)$ will determine the relic density
and scattering cross-section of dark matter. Our primary strategy will be to constrain these parameters by requiring that 
the relic dark matter abundance is regulated by the annihilation at freeze-out either of DM itself or of its mediators,
and then to explore the interplay between the existing constraints and future sensitivity in direct detection and the 
invisible decay width $\Gamma(h\rightarrow 2{\rm DM})$.  To assess the importance of this decay channel, we will 
characterize the invisible Higgs branching with the following figure  of merit \cite{BPtV},
which approximates the dilution of all visible Higgs decay modes in the low $m_h$ regime,
\ba
R_{\rm vis} &=& \frac{\Gamma(h\rightarrow b\bar{b})}{\Gamma(h\rightarrow 2{\rm DM})+\Gamma(h\rightarrow b\bar{b})} \nonumber\\
 &\sim&  \frac{Y_b^2}{Y_b^2 + \fr23 \lambda_{\rm min}^2 v^2/m_h^2}.
\ea
In this expression, $v$ is the electroweak v.e.v.,  the Yukawa coupling of the $b$-quark is 
normalized at the $m_h$ scale, and the phase space factors are neglected.  As the DM mass is  taken 
below $m_{\rm DM} \sim 40$ GeV, the invisible Higgs decay channel becomes dominant. Within background-dominated LHC Higgs searches, 
a detection of the Higgs boson via its conventional decay modes would then require increasing the size of the Higgs dataset by  
at least a factor of $1/R_{\rm vis}^2$.

In the next section, we outline a series of specific model scenarios falling within the general Higgs-mediated 
setting of (\ref{portal}) and (\ref{hidden}), focussing on the link between the invisible Higgs width and the 
direct detection sensitivity.

\section{Scenarios and signatures}

The Higgs portal allows for a number of Higgs-mediated dark matter scenarios, where the set of induced $h-{\rm DM}$ couplings
determines both the direct scattering cross-section and the invisible Higgs width.  Below, we detail a series of modules (or simplified 
models \cite{sm}) which encode the basic physics. Many of these modules can be embedded as part of more comprehensive 
UV theories.

\subsection{WIMPs and the pseudoscalar Higgs portal}

The WIMP scenario of  fermionic DM mediated by the Higgs portal has been discussed 
before (see {\em e.g.} \cite{BKP,Kim,PRV}), focussing on its CP-conserving version (although see \cite{nt}). Here we consider 
a CP-odd combination of the trilinear Higgs portal with a pseudoscalar coupling,
\be
{\cal L} = (H^\dagger H) (AS + \lambda S^2) + \beta S \bar{N} i\gamma_5 N,
\ee
which, on integrating out the heavier scalar $S$ and taking the unimportant coupling $\lambda$ to be small,
leads to
\be
 {\cal L}_{\rm eff} = \lambda_{h} h\bar{N} i\gamma_5 N. \label{CPodd}
\ee
The effective Higgs coupling $\lambda_{h}$ results from 
$S-h$ mixing induced by the $ASH^\dagger H$ term in the Lagrangian, and is 
taken to satisfy the freeze-out condition,
\be
\langle \sigma v \rangle_{\bar{N}N\to SM}  \simeq \fr{3\lambda_h^2}{4\pi}~
\left(\fr{m_b}{m_h}\right)^2~\fr{m_N^2}{m_h^4} \sim 1~{\rm pb}.
\ee
This requires,
\be
\lambda_h^2 \sim 10 \times 
\left(\fr{20~{\rm GeV}}{m_N}\right)^2,
\ee
where we have taken $m_N \sim 20~{\rm GeV}$, which is close to the lower bound given
by the perturbative threshold, $\lambda_h^2 \sim 4\pi$. With $m_h \sim {\cal O}(120)$~GeV, this scenario 
has a limited range for the DM mass, where $\lambda_h$ is always significantly larger than 
the $b$-quark Yukawa coupling, leading to
$R_{\rm vis} \ll 1$ and the possibility of an $O(10^3)$ suppression of all visible Higgs decay modes.

Turning to direct detection (see Fig.~1(a)), we observe that the pseudoscalar 
density $\bar{N} i\gamma_5 N$ vanishes in the nonrelativistic limit, which suppresses the 
elastic DM-nucleon scattering  cross section by an additional factor of $(v/c)^2\sim 10^{-6}$,  
\ba
\sigma^{\rm eq}_{p} &\simeq& \frac{1}{2\pi}  (v/c)^2\times \fr{g_{hpp}^2\lambda_h^2m_p^2}{m_h^4} 
\times \left(\fr{Am_p}{Am_p+m_N}\right)^2 \nonumber\\
 &\la& 10^{-48}~{\rm cm}^2\times \lambda_h^2.
\ea
Note that (distinct from $\si_{\rm SI}$) this is an {\em equivalent} DM-nucleon scattering cross section 
derived from  the DM-nucleus cross section with $A$ nucleons. One observes that not only is 
this cross section well below the current level of direct detection sensitivity, but it may be below the
potentially irreducible background due to the solar neutrino recoil signal \cite{FM}. 

At this point, we should try to address the question of how natural it is to have a dominant CP-odd 
coupling for DM, given the fact that CP violation is small (or flavor-suppressed) in the observable sector 
of the SM. Unfortunately this question does not have a clear-cut answer in the model (\ref{CPodd})
due to the super-renormalizable nature of the portal $ASH^\dagger H$. Indeed, if $S$ is a  pseudoscalar in the dark 
sector, the $S\bar{N} i\gamma_5 N$ coupling conserves all discrete symmetries. With $A$ the CP-violating coupling, 
even a value $A\sim O(M_W)$ may be `small' in the sense that $A \ll \Lambda_{UV}$ given that
 the UV cutoff of this theory is unknown. Changing the CP charge assignment by taking $S$ to be a scalar, we see that
 the source of CP-violation, now shifted to $\beta$, is also well sequestered from any visible
 sector observables due to the need for Higgs mediation. Thus  a fermionic WIMP interacting
 via this pseudoscalar Higgs portal is a viable possibility, and naturally leads 
 to a large invisible Higgs width, $R_{\rm vis}\ll 1$, while having suppressed signatures for
 direct detection.
 We note in passing that were such a scenario realized, there would be an 
observable indirect signature through DM annihilation within overdense regions in the
galactic halo \cite{Fermi} (see {\it e.g.} \cite{indirect} for analyses in the minimal model).

\begin{figure}
\centerline{\includegraphics[viewport=150 356 444 720, clip=true, width=7cm]{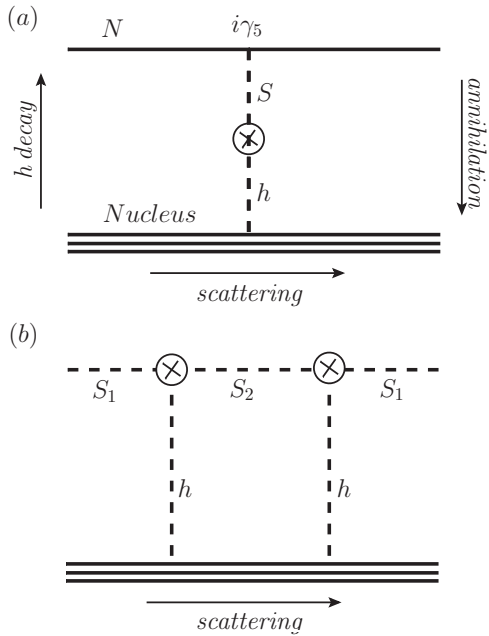}}
 \caption{\footnotesize Schematic illustration of the scattering, annihilation and Higgs decay 
 processes for: (a) fermionic WIMP interactions mediated via the pseudoscalar portal; and (b) split scalar WIMPs, showing 
 loop-level elastic scattering which can dominate over tree-level exchange for $\lambda_{12}\gg \lambda_{11}$.}
\label{f1} 
\end{figure}

\subsection{Split WIMPs}

Multi-component WIMPs with a small splitting between the states have been examined in some detail
in recent years, primarily in connection with inelastic DM models \cite{inelastic}. Here we consider two real scalar 
WIMPs coupled through the Higgs portal,
\be
{\cal L} = \sum_{i,j = 1,2}S_i\left(\Lambda_{ij} (H^\dagger H) -\fr12M^2_{ij}\right)S_j,
\ee
where $\Lambda$ and $M^2$ are the 2$\times$2 real symmetric matrices. DM stability requires a suitable
dark symmetry, which forbids the $A$ term couplings. A $\mathbb{Z}_2$ acting as $S_i \rightarrow -S_i$
is one minimal option, among many alternatives.
After electroweak symmetry breaking and without the loss of generality we can choose the mass terms to be 
diagonal and write the effective Higgs-DM Lagrangian as 
\ba
{\cal L} _{\rm eff} &=& -\fr12(m_1^2S_1^2 + m_2^2S_2^2) \nonumber\\
 && + h(\lambda_{11}S_1^2 +2 \lambda_{12}S_1S_2 + 
\lambda_{22}S_2^2),
\ea
where we also assume $m_2>m_1$. In the present note,  we will not attempt to analyze 
the full parameter space for $m_1,~m_2$ and $\lambda_{ij}$. Instead, we will
concentrate on the case where the mass splitting is relatively small,  
\be
 \De m = m_2 - m_1\la 0.1 m_1, \label{hier1}
\ee
 and the couplings exhibit the hierarchical pattern, 
\be
\lambda_{22}\gg \lambda_{12} , \lambda_{11}.
\label{hier2}
\ee 
Choosing this pattern of couplings, which we will justify below, ensures that the model has the following properties: 
\begin{itemize}
\item $S_1$ is a stable DM candidate, while $S_2$ is unstable, $S_2 \to S_1 +{\rm SM}$. 

\item The cosmological abundance of $S_1$ is controlled via coannihilation: 
$S_1+{\rm SM}\to S_2+{\rm SM}$ followed by $S_2+S_2\to {\rm SM}$.

\item The elastic scattering cross section of $S_1$ on nucleons is suppressed relative to the minimal model by 
$(\lambda_{11}/\lambda_{22})^2$ (or loop suppressed if $\lambda_{11}$ is sufficiently small \cite{bpr}, as exhibited in Fig.~1(b)). If the 
mass splitting $\De m$ is in the keV range or below, an inelastic 
component to scattering is also present but  suppressed by $(\lambda_{12}/\lambda_{22})^2$. 

\item For DM masses below roughly 40 GeV, the Higgs decay is totally dominated by $h\to2S_2$. 
Depending on the size of $\lambda_{12}$, the subsequent decay $S_2\to S_1+{\rm SM}$ may or may not happen within the detector 
volume,  resulting in either a `buried' or an invisible Higgs decay signature. 
\end{itemize}

We can estimate the the size of the off-diagonal coupling $\lambda_{12}$ needed to ensure that $S_1$ and $S_2$ 
stay chemically coupled at freeze-out by comparing the rate of Higgs-induced $1\leftrightarrow 2$ conversion, 
$S_1 + SM \to S_2 +SM$, with the Hubble rate at $T\sim 0.05 m_1$. Estimating the scattering of $m_S\sim 20$ GeV 
$S$-particles on charm quarks at $T\sim 1$ GeV, we arrive at the condition
\be
\lambda_{12}^2 \ga 10^{-6}.
\label{coan}
\ee

We now address the naturalness of  the hierarchy (\ref{hier1},\ref{hier2}).  A simple scenario which achieves it assumes 
that initially the matrix $\Lambda$ is dominated by one entry, $\Lambda \simeq {\rm diag}(0, \lambda_{22})$, and the mass matrix is also nearly diagonal 
with a small off-diagonal entry, $m^2_{12}\ll m_2^2-m_1^2$. In this case, the mixing angle required to go to the mass 
eigenstate basis is small, $\theta \sim m_{12}^2/(m_2^2-m_1^2)$, and the induced $hS_1S_2$ and $hS_1^2$ 
couplings are suppressed: $\lambda_{12} \sim \theta \lambda_{22}$; $\lambda_{11} \sim \theta^2 \lambda_{22}$. 
The small size of $m_{12}^2$ relative to $m_2^2-m_1^2$ can arise naturally if there are separate 
approximate discrete symmetries for $S_1$ and $S_2$ broken by this term. We conclude that this scenario is 
viable, and does not require an excessive tuning of parameters. In view of the coannihilation requirement (\ref{coan}),
$\theta$ can  be as small as $10^{-3}$, in which case the direct detection cross sections are suppressed by 
a factor of $O(\theta^4) \sim 10^{-12}$.

\subsection{Secluded WIMPs}

The secluded regime \cite{PRV,FW} makes use of the Lagrangian 
\be
{\cal L} =  H^\dagger H ( AS + \lambda S^2) + \al S \bar{N} N,
\ee
and requires that $m_N > m_S$, so that annihilation proceeds via $\bar{N}N\rightarrow SS$, 
with $S$ decaying on-shell to the SM. Requiring $N$ to be a thermal WIMP leads to the
usual restriction that  $\alpha^4/m_N^2$ is fixed to be $O({\rm pb})$.
The Higgs can decay directly to the mediators $h \rightarrow SS$, with a width controlled by the 
coupling $\lambda$ which does not affect the abundance of $N$. Taking $\lambda$ above 
$Y_b$ ensures $R_{\rm vis} <1$, so that the Higgs width to $2S$ can be sizeable. 
 As in the split WIMP scenario, this  decay will be invisible if $S$ is sufficiently long-lived to 
escape the detector, while it will be `buried' if $S$ decays occur inside the detector, leading to
multiple jets  in the final state. The WIMP-nucleon scattering cross section can be made almost 
arbitrarily small \cite{PRV}, and in practice taking $A/v \sim 10^{-3}$ renders the scattering rate 
below the neutrino background for direct detection. On the other hand, this model does have 
indirect DM detection signatures, which can be enhanced in the case of nonrelativistic annihilation
provided that $m_S$ is light \cite{FW,mrwch,AFSW,PR}.

\subsection{Super-WIMPs}

A particularly  `signature-poor' variant of the scenarios considered here comprises a
neutral particle $N$ sufficiently weakly coupled to the observable sector that throughout the thermal history of the 
Universe it remains sparsely populated compared to other species, i.e. a super-WIMP (see {\it e.g} \cite{ens}). The Lagrangian 
can be taken in a form that closely resembles the minimal scalar DM model, 
\be
 {\cal L} = \lambda (H^\dagger H) S^2 + \al S \bar{N} N - V(S),
\ee
where the mediator $S$ is relatively heavy, $m_S>2m_N$.
The thermal history consists of the normal annihilation of $S$ at the freeze-out, 
followed by the late decay $S\to2N$ producing the relic dark matter population.
Assuming that the coupling constant $\alpha$ satisfies the criterion,
\be 
 \alpha^2 \la 10^{-24},
\ee
the direct thermal production of $N$ states (e.g. via $Sh \rightarrow N \bar{N}$) is subdominant to $S$ decays and $N$ is
a super-WIMP. The coupling $\lambda$ required to ensure the appropriate 
relic abundance of $S$ (and thus $N$) can then be obtained from 
the corresponding coupling in the minimal scalar model by the following rescaling,
\be
\lambda = \fr{2m_N}{m_S}\times \lambda_{\rm min}.
\ee

The only signature of this model is the enhanced $h\to2S$ decay that can easily dominate over the SM 
channels if $2m_N/m_S$ is not too small. Unlike other super-WIMP models (e.g. NLSP-to-gravitino decays), the decay 
of $S\to 2N$ occurs completely within the dark sector and does not have additional BBN/CMB-related signatures \cite{ens}. 

\subsection{WIMPs and the SUSY Higgs portal}

Appropriately mixed $\tilde{B}/\tilde{H}$ neutralino LSP 
candidates in the MSSM are archetypal examples of WIMPs which can undergo Higgs-mediated 
elastic scattering at the current direct-detection threshold (see e.g. \cite{bfs} for an analysis 
in the low mass range). However, this example does not fit within the scenarios outlined above, primarily  
because it relies on  $\tan\beta = \langle H_2\rangle /\langle H_1\rangle $ being large, which 
enhances DM annihilation mediated  by the pseudoscalar Higgs $A$, and  neutralino-nucleon 
scattering mediated by $H$ exchange. The invisible width of the lightest Higgs boson $h$ decaying
to neutralinos is suppressed by $m_\chi^2/\mu^2$ due to the small mixing of 
$\tilde{B}$ with $\tilde{H}_2$. Models with light neutralino DM also 
typically have a number of charged states (Higgses $H^\pm$, sfermions etc.) 
near the weak scale, which allows for discovery via channels unrelated to 
DM. However, turning to NMSSM scenarios, the chances for invisible or hidden Higgs decays are substantially 
higher \cite{Gunion}. 

Here we extend the MSSM particle content by singlet chiral superfields $N$ and $S$, in close 
analogy with Eqs.~(\ref{portal}) and (\ref{hidden}), while requiring all superpartners of the SM fields to be heavy. 
An example of this NMSSM-type extension, with a supersymmetric Higgs portal, is given 
by the superpotential,
\be
 W =  \lambda_S H_1H_2 S + m_SS^2+ (M + \al S)N^2.
\ee
If, in addition to SM superpartners, $H$, $A$ and $H^\pm$  are heavy and $\tan \beta \gg 1$, this model 
reduces to the SM (with $H_2$ being the SM-like Higgs doublet)  plus the SUSY multiplets of $S$ and $N$. 
By varying the couplings, one can find regimes reproducing most of the SM Higgs portal models discussed above. 
The scalar potential contains  the terms $V = |\mu H_2+\lambda_S H_2 S|^2$, from which we can identify 
the couplings to the complex scalar $S$ in (\ref{portal}) as $A=\mu\lambda_S$ and $\lambda=|\lambda_S|^2$. 
Choosing these couplings appropriately, we reproduce the super-WIMP and secluded WIMP models
with states from the $N$ multiplet playing the role of DM.
The pseudoscalar Higgs portal can be constructed by taking $\al$ real and choosing arg$(\mu^*\lambda_S)=\pm\pi/2$.
In this case, only Im$(S)$ couples the fermionic DM candidate $N$ to the Higgs portal via the $\bar N i \gamma_5 N$ bilinear, 
while Re$(S)$ will not couple to the Higgs portal at all. Finally, split WIMPs can be obtained by introducing multiple $N_i$ states 
with small mass splittings, while allowing just one to have a large coupling to the $S$ mediator field. 
In all these models, the lightest SUSY Higgs state $h$ can have a significant (or dominant) invisible 
branching fraction directly to DM states and/or its mediators, while 
the direct detection cross sections are suppressed.

\section{Concluding Remarks}

The simplicity of the varied Higgs portal scenarios considered above serves to underscore the point
that generic models of Higgs-mediated dark matter -- beyond the minimal model of scalar DM -- do not imply a
rigid link between the invisible Higgs decay width and the DM direct detection signal. Thus the absence
of a signal in direct detection need not preclude a sizeable invisible Higgs width even if dark matter is
predominantly Higgs-mediated. This emphasizes the important role that invisible Higgs searches
could play in the eventuality that conventional Higgs signatures are found to be suppressed and/or excluded at the 
level of SM cross sections and decay rates. 

Finally we note that scenarios with a large invisible Higgs branching may have interesting cosmological 
implications, ranging from changes to the thermal history of the SM and dark sectors (see {\it e.g.} \cite{thermal}), to 
modifications of the electroweak phase transition \cite{ekr} which may impact scenarios for electroweak baryogenesis.

\section*{Acknowledgements}

We would like to thank B. Batell and D. McKeen for discussions.  The work of  M.P. and A.R. is supported 
in part by NSERC, Canada, and research at the Perimeter Institute is supported in part by the Government 
of Canada through NSERC and by the Province of Ontario through MEDT.

\end{document}